\documentclass[letterpaper,english,aps,letterpaper,english,reprint]{revtex4-1}
\usepackage{lmodern}
\usepackage{lmodern}
\usepackage[latin9]{inputenc}
\usepackage{textcomp}
\usepackage{amsmath}
\usepackage{amssymb}
\usepackage{graphicx}
\usepackage{wasysym}
\usepackage{esint}

\makeatletter

\newcommand*\LyXThinSpace{\,\hspace{0pt}}

\usepackage{mathrsfs}

\usepackage{babel}

\makeatother

\usepackage{babel}
\begin{document}

\title{Goos-H%
\mbox{%
ä%
}nchen shift of a spin-wave beam transmitted through anisotropic interface
between two ferromagnets}

\author{P. Gruszecki,$^{1}$ M. Mailyan,$^{2}$ O. Gorobets,$^{2,3}$ and
M. Krawczyk$^{1}$}

\affiliation{$^{1}$Faculty of Physics, Adam Mickiewicz University in Poznan,
Umultowska 85, Pozna\'{n}, 61-614, Poland~~\\
 $^{2}$ National Technical University of Ukraine \textquotedblright Kyiv
Polytechnic Institute\textquotedblright , 37 Peremogy ave., 03056,
Kyiv, Ukraine~~\\
 $^{3}$Institute of Magnetism, National Academy of Sciences of Ukraine,
36-b Vernadskogo st., 03142, Kyiv, Ukraine}

\date{\today}
\begin{abstract}
The main object of investigation in magnonics, spin waves (SWs) are
promising information carriers. Presently the most commonly studied
are plane wave-like SWs and SWs propagating in confined structures,
such as waveguides. Here we consider a Gaussian SW beam obliquely
incident on an ultra-narrow interface between two identical ferromagnetic
materials. We use an analytical model and micromagnetic simulations
for an in-depth analysis of the influence of the interface properties,
in particular the magnetic anisotropy, on the transmission of the
SW beam. We derive analytical formulas for the reflectance, transmittance,
phase shift and Goos-Hänchen (GH) shift for beams reflected and refracted
by an interface between two semi-infinite ferromagnetic media; the
results for the refracted beam are the first to be reported to date.
The GH shifts in SW beam reflection and transmission are confirmed
by micromagnetic simulations in the thin-film geometry. We demonstrate
the dependence of the characteristic properties\inputencoding{latin1}{
}\inputencoding{latin9}on the magnetic anisotropy at the interface,
the angle of incidence and the frequency of the SWs. We also propose
a method for the excitation of high-quality SW beams in micromagnetic
simulations. 
\end{abstract}
\maketitle

\section{\label{sec:level1}Introduction\label{sec:Introduction}}

Moore's law in its basic form postulated in 1965~\citep{More65},
stating that the number of transistors in a dense integrated circuit
(or chip's doubles approximately every two years, is now nearing its
end~\citep{Waldrop16}. In the light of the still increasing demand
for computational resources, the presently dominating CMOS circuits
are reaching their limits in terms of miniaturization, performance
and energy consumption~\citep{Waldrop16,Lloyd00}. Moreover, the
costs are rising while the benefits of further miniaturization are
decreasing~\citep{Waldrop16}. This brings about the necessity to
create a new class of devices with enhanced performance and functionality
for various applications to supplement or even replace CMOS circuits~\citep{Bernstein10,Nikonov13}.
Spin waves (SWs) are among the potential candidates for replacement
of electric charges as information carriers~\citep{Krawczyk14,Chumak14,Ed2015}.

The main object of investigation in the emerging branch of modern
magnetism known as magnonics~\citep{Kruglyak10,Demokritov13}, SWs
are magnetization excitations propagating without charge transport.
This excludes Joule heating and implies that the application of SWs
in computing devices could significantly reduce the energy consumption
with respect to the charge-based alternatives \citep{Nikonov13,Chumak2015b}.
Spin waves have frequencies ranging from several~GHz to hundreds
of~GHz with wavelengths 4 to 5 orders of magnitude shorter than those
of electromagnetic waves of the same frequencies. Moreover, even in
homogeneous planar structures SWs have a nontrivial dispersion relation,
resulting in an exceptional richness of potential properties that
could be used for SW manipulation; these properties have no counterpart
in photonics and electronics~\citep{Krawczyk14}.

Potentially, SWs guarantee high operating frequencies and low energy
consumption with preserved CMOS level of miniaturization and possible
integration with present CMOS circuits. Solutions known from photonics,
spintronics and electronics can be applied in magnonics as well. Furthermore,
the use of SWs as information carriers would pave the way to wave-based
non-Boolean computing~\citep{Csaba14}, holographic memory \citep{Gertz16,Kozhevnikov15}
or the physical realization of neural networks \citep{Locatelli2014}.
Crucial in these approaches is the manipulation of both amplitude
and phase of the SWs. In magnonics this can be realized by means of
magnonic crystals or SW scattering by disturbances such as holes \citep{Gertz16},
domain walls \citep{Macke2010} or magnetic elements placed on waveguide
cross junctions \citep{Kozhevnikov15}. Another possible approach,
which has emerged in modern photonics for electromagnetic waves, is
based on the use of materials referred to as metasurfaces for wave
manipulation at sub-wavelength distances~\citep{Yu2014}.

The reflection and refraction of SWs are determined by the magnetic
properties of the ferromagnetic media and by the interface boundary
conditions. In the theoretical and experimental studies of spin-wave
reflection \citep{Chumak09,Kim08} and refraction \citep{Kim08,Klos15,Stigloher16}
reported to date SWs are mostly treated as plane waves. %OS: considered in the plane-wave approach? 
Other kinds of excitations, specifically, coherent low-divergence
spin-wave beams (SWBs), the application of which would open new possibilities,
have not been explored to date. There are only a few theoretical and
experimental studies on the formation of low-frequency SW beams; research
in this field is hampered by caustics, nonlinear effects and difficulties
related to excitation by nanooscillators or width-modulated microwave
transducers \citep{Vugal88,Floyd84,Schneider2010,Gieniusz13,Kostylev2011,Boyle96,Bauer97,Houshang2015,Gruszecki15-antennas}.

From the theoretical point of view the study of SW reflection and
refraction can be regarded as the investigation of the amplitude and
phase changes (in relation to the incident SWs) that the reflected
and transmitted SWs, respectively, undergo at the interface. Very
convenient parameters in that study are reflectance~$R$ (the power
ratio of the reflected and incident SWs) and transmittance~$T$ (the
power ratio of the transmitted and incident SWs), and the phase shifts~$\Delta\varphi_{\mathrm{r}}$
and $\Delta\varphi_{\mathrm{t}}$ of the reflected and transmitted
waves, respectively, in relation to the incident SWs. It is noteworthy
that one of the physical consequences of those phase shifts can be
the occurrence of a lateral shift~$\Delta_{\mathrm{r}}$ or $\Delta_{\mathrm{t}}$
of the beam spots along the interface in the respective effects. Known
from optics and first demonstrated for reflected light in 1947 by
H.~Hänchen and F.~Goos~\citep{Hanchen47}, this wave phenomenon
is referred to as the Goos-Hänchen effect~(GHE). It occurs also in
acoustics, where it is known as the Schoch displacement~\citep{Declercq2008},
in electronics~\citep{Chen2013} and in neutron waves~\citep{Haan10}.

Although shown theoretically not to occur in magnetostatic waves (i.e.,
SWs in the limit of negligible exchange interaction~\citep{Yasumoto83}),
the GHE has been demonstrated, also theoretically, in reflection of
purely exchange SWs (i.e., high-frequency SWs with neglected dipole-dipole
interactions) by an interface between two semi-infinite ferromagnetic
films \citep{Dadoenkova2012}. The effect was later confirmed by micromagnetic
simulations for exchange-dipolar SWs reflected by the edge of a magnetic
thin film~\citep{Gruszecki14}. The magnetic properties at the film
edge have been shown to be crucial for the lateral shift of the SW
beam; specifically, the value of~$\Delta_{\mathrm{r}}$ is very sensitive
even to slight changes in the magnetic surface anisotropy at the film
edge~\citep{Gruszecki15}.

Full elucidation of SW reflection and refraction and the influence
of the boundary conditions on the reflectance, transmittance and phase
shifts would open the possibility of manipulating not only the amplitudes,
but also the phases of SWs at very short distances. This could be
critical for the future development of magnonic devices.

In this paper we analyze theoretically and numerically SW refraction
by an ultra-narrow interface between two ferromagnetic materials.
We consider high-frequency SWs and study the influence of the interface
on their reflection and, especially, refraction. In particular, we
investigate the role of the magnetic anisotropy introduced at the
interface and focus on the transmittance and the GH shifts of the
transmitted SW beams. We propose an analytical model for purely exchange
SWs in two semi-infinite materials, and derive formulas for the reflectance
and transmittance and the respective GH shifts.

These results are verified by micromagnetic simulations (MSs) of a
SW beam passing through an interface between two semi-infinite ferromagnetic
thin films. The performed MSs take also account of the dipolar interaction.
The data obtained demonstrate that the reflection and transmission
of SWs are sensitive to even slight changes in the magnetic anisotropy
introduced at the interface, the thickness of which is much smaller
than the wavelength of the SWs. The transmittance is shown to decrease
and the GH shift to grow with decreasing SW frequency and angle of
incidence.

Our findings can be of interest and use wherever SW phase and amplitude
manipulation is required, including logic and microwave applications.
Moreover, we believe that the control of SWs at sub-wavelength distances
can initiate a new research direction, which, by analogy with photonics,
can be called magnonic metasurfaces. We have also developed an efficient
method for SWB excitation in MSs, which can be easily exploited in
magnonic studies.

The paper is organized as follows. In Sec.~\ref{sec:Methods} we
present the analytical model of SW reflection and refraction, and
the micromagnetic simulations. The obtained analytical results and
simulation data are discussed in Sec.~\ref{sec:Results}. Conclusions
are provided in Sec.~\ref{sec:Discussion}. The derived analytical
formulas for the GH shifts are presented in Appendix~A, while Appendix~B
describes the method of SWB excitation used in the MSs.

\section{Methods\label{sec:Methods}}

\subsection{Analytical model\label{subsec:Methods_Analitical-model}}

\begin{figure}
\includegraphics[width=8.6cm]{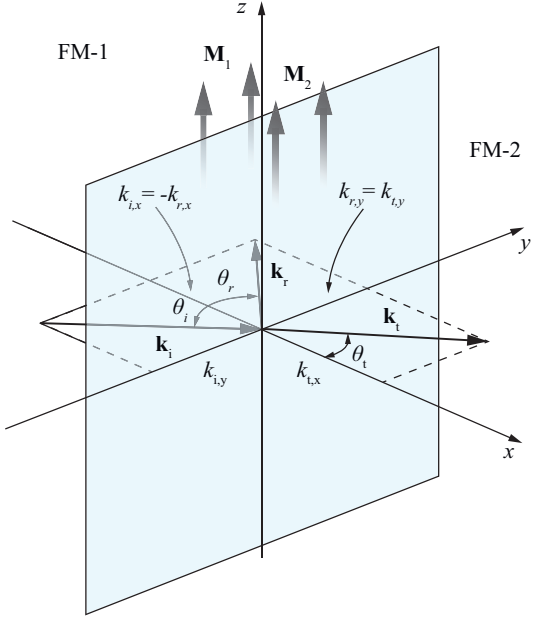} \protect\caption{Schematic representation of the system used in the analytical model.
Two semi-infinite ferromagnetic materials, FM1 and FM2, are separated
by an interface (gray plane) lying in the $yOz$ plane. The external
magnetic field, static magnetization and anisotropy field are all
oriented along the $z$-axis. The plane of incidence is the $xOy$
plane. \label{fig:FigAM_schema}}
\end{figure}

Let us begin by constructing an analytical model of SW reflection
and refraction by an interface, extending along the $y$- and $z$-axes,
between two semi-infinite ferromagnetic materials FM\nobreakdash-1
and FM\nobreakdash-2 (Fig.~\ref{fig:FigAM_schema}). We assume uniform
magnetization of the system by an external magnetic field $\mathbf{H}=\left[0,0,H_{0}\right]$
parallel to the interface ($yOz$ plane) and perpendicular to the
plane of incidence ($xOy$ plane). In both materials we consider the
same direction of magnetocrystalline anisotropy, parallel to that
of the surface magnetic anisotropy, along the $z$-axis: $\mathbf{n}^{(1)}=\mathbf{n}^{(2)}=\mathbf{e}_{z}$.
We will only analyze here high-frequency short-wavelength SWs (i.e.,
exchange SWs), in which case the influence of the dipolar interaction
can be neglected. The considered SWs are also assumed to be uniform
along the $z$-axis. %\includegraphics[width=8.6cm]{\string"Spin wave propagation\string".jpg}

The total energy of the system composed of two ferromagnetic materials
(indicated with superscripts~$^{(1)}$ and~$^{(2)}$) can be written
as: 
\begin{equation}
\begin{array}{cc}
W= & \underset{V}{\int}dv\left[w_{\mathrm{H}}^{\left(1\right)}+w_{\mathrm{H}}^{\left(2\right)}+w_{\mathrm{ex}}^{\left(1\right)}+w_{\mathrm{ex}}^{\left(2\right)}\right.\\
 & \left.+w_{\mathrm{anis}}^{\left(1\right)}+w_{\mathrm{anis}}^{\left(2\right)}+w_{\mathrm{ex}}^{\left(12\right)}+w_{\mathrm{anis}}^{\left(12\right)}\right],
\end{array}\label{Eq:Total_energy}
\end{equation}
where the integration runs over the whole volume.

The term $w_{\mathrm{H}}^{\left(l\right)}=-\mathbf{M}^{(l)}\cdot\mathbf{H}$
is the Zeeman energy density, where $\mathbf{\mathbf{M}}^{(l)}=\left[m_{x}^{(l)},m_{y,}^{(l)}M_{l}\right]$
denotes the magnetization vector in the $l$-th ferromagnet and $M_{l}$
is its saturation magnetization.

The next two terms, $w_{\mathrm{ex}}^{\left(l\right)}=\dfrac{1}{2}\alpha_{l}(x)\left(\nabla\mathbf{M}^{(l)}\right)^{2}$,
are the exchange energy densities; $\alpha_{l}=A_{l}/M_{l}^{2}$ denotes
the exchange length and $A_{l}$ is the exchange constant. The term
$w_{\mathrm{anis}}^{\left(l\right)}=-\dfrac{1}{2}\beta_{l}(x)\left(\mathbf{M}^{(l)}\cdot\mathbf{n}^{(l)}\right)^{2}$
is the magnetic anisotropy energy density in the $l$-th ferromagnet;
$\beta_{l}=K_{l}/M_{l}^{2}$, where $K_{l}$ is the uniaxial magnetic
anisotropy constant, and $\mathbf{n}^{(l)}$ is a unit vector pointing
in the direction of the easy axis in the $l$-th ferromagnet.

The term $w_{\mathrm{ex}}^{\left(12\right)}=A_{12}\mathbf{M}^{(1)}\cdot\mathbf{M}^{(2)}\Theta_{\mathrm{H}}(x)\Theta_{\mathrm{H}}(-x+\delta)$
is the energy density of interlayer exchange coupling at the interface
between the ferromagnets; $\Theta_{\mathrm{H}}(x)$ is the Heaviside
step function, $\delta$ the width of the interface, and $A_{12}$
a parameter of uniform exchange interaction; $A_{12}=\xi A_{\mathrm{int,S}}/M_{\mathrm{int}}^{2}$,
where $A_{\mathrm{int,S}}$ denotes the effective surface exchange
constant of the interface ($A_{\mathrm{int,S}}=A_{\mathrm{int}}/\delta$,
with $A_{\mathrm{int}}$ being the exchange constant of the interface
region), $M_{\mathrm{int}}$ is the saturation magnetization of the
interface, and $\xi$ a proportionality coefficient (we assume $\xi=400$).

The last term in Eq.~(\ref{Eq:Total_energy}), $w_{\mathrm{anis}}^{\left(12\right)}=-\dfrac{1}{2}\left[\beta_{12}\left(\mathbf{M}^{(1)}\cdot\mathbf{n}^{(1)}\right)\left(\mathbf{M}^{(2)}\cdot\mathbf{n}^{(2)}\right)\right]\Theta_{\mathrm{H}}(x)\Theta_{\mathrm{H}}(-x+\delta)$,
is the energy density of surface magnetic anisotropy at the interface;
$\beta_{12}=K_{12}/M_{\mathrm{int}}^{2}$ is an anisotropy parameter,
with $K_{12}$ denoting the uniaxial magnetic anisotropy constant
at the interface (which can be regarded as stemming from the surface
magnetic anisotropy, $K_{12}\equiv K_{\mathrm{S}}/\delta$).

The SW dynamics in this system can be described by the Landau-Lifshitz
(LL) equations for both ferromagnetic materials: 
\begin{equation}
\begin{cases}
\dfrac{\partial\mathbf{M}^{(1)}}{\partial t}= & \left|\gamma\right|\mathbf{M}^{\left(1\right)}\times\mathbf{H}_{\mathrm{eff}}^{\left(1\right)}\\
\dfrac{\partial\mathbf{M}^{(2)}}{\partial t}= & \left|\gamma\right|\mathbf{M}^{\left(2\right)}\times\mathbf{H}_{\mathrm{eff}}^{\left(2\right)},
\end{cases}\label{eq:LLE_AM-1}
\end{equation}
where $\gamma$ is the gyromagnetic ratio. We will use the linear
approximation, based on the assumption that the dynamic components
of the magnetization are much smaller than the saturation magnetization,
$m_{x,y}^{(l)}\ll M_{l}$, and the latter can be treated as constant.
The effective magnetic field~$\mathbf{H}_{\mathrm{eff}}^{\left(l\right)}$
in each material can be determined as the functional derivative of
the total energy, defined in Eq.~(\ref{Eq:Total_energy}), with respect
to the magnetization vector: 
\begin{equation}
\mathbf{H}_{\mathrm{eff}}^{\left(l\right)}=-\dfrac{\delta W}{\delta\mathbf{M}^{(l)}}=-\dfrac{\partial w}{\partial\mathbf{M}^{(l)}}+\sum_{\zeta\in\{x,y,z\}}\dfrac{\mathrm{d}}{\mathrm{d}\zeta}\dfrac{\partial w}{\partial\left(\dfrac{\mathrm{d}\mathbf{M}}{\mathrm{d}\zeta}\right)},\label{eq:H_eff0}
\end{equation}
where $w$ is the energy density, the integral kernel in Eq.~(\ref{Eq:Total_energy}).

Assuming plane-wave solutions of the LL equations~(\ref{eq:LLE_AM-1}),
\mbox{%
$m_{x,y}^{(l)}\propto\text{exp}[i(\mathbf{k}_{l}\cdot\mathbf{r}-\omega_{l}t)]$%
}, in each of the ferromagnetic materials, we obtain the SW dispersion
relation: 
\begin{equation}
\omega_{l}\left(\mathbf{k}_{l}\right)=\left|\gamma\right|\left(H_{0}+\beta_{l}M_{l}+M_{l}\alpha_{l}k_{l}^{2}\right),\label{Eq:Dispersion}
\end{equation}
where $\mathbf{k}_{l}$ is the wavevector and $\omega_{l}$ the angular
frequency of SWs in the $l$-th ferromagnet \citep{Kosevich1983}.

The LL equations~(\ref{eq:LLE_AM-1}) can be integrated over the
interface region in the limit of infinitely narrow interface~\cite{footnote1}:
\begin{equation}
{\displaystyle \varint_{-0}^{+0}}\left[\dfrac{\partial\mathbf{M}^{(l)}}{\partial t}-\left|\gamma\right|\mathbf{M}^{\left(l\right)}\times\mathbf{H}_{\mathrm{eff}}^{\left(l\right)}\right]dx=0.
\end{equation}
The integration of the above equations (for $l=1$ and $l=2$) with
the effective magnetic field as expressed in~(\ref{eq:H_eff0}) yields
the boundary conditions in the form of a set of equations linking
the dynamic components of the magnetization on both sides of the interface:
\begin{equation}
\left\{ \begin{array}{l}
\left.\left(A_{12}m_{n}^{(2)}+Dm_{n}^{(1)}+\alpha_{1}\dfrac{\partial m_{n}^{(1)}}{\partial x}\right)\right|_{x=-0}=0\\
\\
\left.\left(A_{12}m_{n}^{(1)}+Cm_{n}^{(2)}-\alpha_{2}\dfrac{\partial m_{n}^{(2)}}{\partial x}\right)\right|_{x=+0}=0,
\end{array}\right.\label{eq:BC_AM}
\end{equation}
where $n=x,y$, $D=-\left[\left(A_{12}-\beta_{12}\right)\zeta-\beta_{1}\right]\delta$,
$C=-\left[\left(A_{12}-\beta_{12}\right)/\zeta+\beta_{2}\right]\delta$,
and $\zeta=M_{2}/M_{1}$. The physical meaning of the parameters $D$
and $C$ is that of effective values obtained by averaging the finite
width~$\delta$ over the interface region (see Ref.~\citep{Kruglyak2014},
Section~4).

Having the boundary conditions~(\ref{eq:BC_AM}), we can derive the
Fresnel equations for exchange SWs. To this end we describe incident
and reflected circularly polarized SWs in FM-1 and refracted SWs in
FM-2 as monochromatic plane waves: 
\begin{equation}
\begin{cases}
\left(m_{x}^{(1)}+im_{y}^{(1)}\right)= & \mathrm{e}^{i\left(\mathbf{\mathbf{k}_{\mathrm{i}}}\cdot\mathbf{r}-\omega t\right)}+r\mathrm{e}^{i\left(\mathbf{k}_{\mathrm{r}}\cdot\mathbf{r}-\omega t+\varphi_{\mathrm{r}}\right)}\\
\left(m_{x}^{(2)}+im_{y}^{(2)}\right)= & t\mathrm{e}^{i\left(\mathbf{k}_{\mathrm{t}}\cdot\mathbf{r}-\omega t+\varphi_{\mathrm{t}}\right)},
\end{cases}\label{eq:PlaneWaves}
\end{equation}
where $r$ and $t$ are the reflection and transmission coefficients,
respectively. A SW incident at an angle~$\theta_{\mathrm{i}}$ has
a wavevector $\mathbf{k}_{\mathrm{i}}=(k_{\mathrm{i},x}\mathbf{e}_{x}+k_{\mathrm{i},y}\mathbf{e}_{y})=k_{\mathrm{i}}(\mathbf{e}_{x}\cos\theta_{\mathrm{i}}+\mathbf{e}_{y}\sin\theta_{\mathrm{i}})$.
Similarly, the reflected and transmitted SWs have wavevectors $\mathbf{k}_{\mathrm{r}}=(k_{\mathrm{r},x}\mathbf{e}_{x}+k_{\mathrm{r},y}\mathbf{e}_{y})$
and $\mathbf{k}_{\mathrm{t}}=(k_{\mathrm{t},x}\mathbf{e}_{x}+k_{\mathrm{t},y}\mathbf{e}_{y})$,
respectively. Due to the translational symmetry along the interface
the wavevector component tangential to the interface is conserved,
$k_{\mathrm{i},y}=k_{\mathrm{r},y}=k_{\mathrm{t},y}$. Moreover, the
isotropic dispersion relation implies $k_{\mathrm{i},x}=-k_{\mathrm{r},x}$. 

The substitution of Eq.~(\ref{eq:PlaneWaves}) into the boundary
conditions~(\ref{eq:BC_AM}) yields the Fresnel amplitude coefficients
for reflected and refracted SWs: 
\begin{equation}
r=\sqrt{\dfrac{\left(A_{12}^{2}-CD-\alpha_{1}\alpha_{2}k_{\mathrm{r},x}k_{\mathrm{t},x}\right)^{2}+\left(D\alpha_{2}k_{\mathrm{t},x}-C\alpha_{1}k_{\mathrm{r},x}\right)^{2}}{\left(A_{12}^{2}-CD+\alpha_{1}\alpha_{2}k_{\mathrm{r},x}k_{\mathrm{t},x}\right)^{2}+\left(D\alpha_{2}k_{\mathrm{t},x}+C\alpha_{1}k_{\mathrm{r},x}\right)^{2}}},\label{Eq:r}
\end{equation}
\begin{equation}
t=\dfrac{2A_{12}\alpha_{1}k_{\mathrm{r},x}}{\sqrt{\left(A_{12}^{2}-CD+\alpha_{1}\alpha_{2}k_{\mathrm{r},x}k_{\mathrm{t},x}\right)^{2}+\left(D\alpha_{2}k_{\mathrm{t},x}+C\alpha_{1}k_{\mathrm{r},x}\right)^{2}}}.\label{Eq:t}
\end{equation}

Reflectance $R$ and transmittance $T$ describe how the energy carried
by a SW changes as a result of reflection and refraction, respectively,
in relation to the incident wave. Thus, in addition to their dependence
on the amplitude coefficients, those parameters can also vary with
angle of incidence. In the case of reflection $\theta_{\mathrm{i}}=\theta_{\mathrm{r}}$
and 
\begin{equation}
R=r^{2}.
\end{equation}
In refraction: 
\begin{equation}
T=t^{2}\frac{\cos(\theta_{\mathrm{t}})}{\cos(\theta_{\mathrm{i}})}.\label{Eq:transm}
\end{equation}

We can also determine the phase differences $\varphi_{t}$ and $\varphi_{r}$
between transmitted/reflected and incident SWs. The phase shift of
a reflected SW with respect to the incident wave is: 
\begin{widetext}
\begin{eqnarray}
\varphi_{\mathrm{r}} & = & \mathrm{arcctg}\left(-\dfrac{D\alpha_{2}\sqrt{k_{\mathrm{t}}^{2}-k_{\mathrm{r},y}^{2}}+C\alpha_{1}\sqrt{k_{\mathrm{r}}^{2}-k_{\mathrm{r},y}^{2}}}{A_{12}^{2}-CD+\alpha_{1}\alpha_{2}\sqrt{(k_{\mathrm{r}}^{2}-k_{\mathrm{r},y}^{2})}\sqrt{(k_{\mathrm{t}}^{2}-k_{\mathrm{r},y}^{2})}}\right)\nonumber \\
 & + & \mathrm{arcctg}\left(\dfrac{D\alpha_{2}\sqrt{k_{\mathrm{t}}^{2}-k_{\mathrm{r},y}^{2}}-C\alpha_{1}\sqrt{k_{\mathrm{r}}^{2}-k_{\mathrm{r},y}^{2}}}{A_{12}^{2}-CD-\alpha_{1}\alpha_{2}\sqrt{(k_{\mathrm{r}}^{2}-k_{\mathrm{r},y}^{2})}\sqrt{(k_{\mathrm{t}}^{2}-k_{\mathrm{r},y}^{2})}}\right).\label{Eq:phase_r}
\end{eqnarray}
The phase of a transmitted SW is shifted by: 
\begin{equation}
\varphi_{\mathrm{t}}=\mathrm{arcctg}\left(-\dfrac{D\alpha_{2}\sqrt{k_{\mathrm{t}}^{2}-k_{\mathrm{r},y}^{2}}+C\alpha_{1}\sqrt{k_{\mathrm{r}}^{2}-k_{\mathrm{r},y}^{2}}}{A_{12}^{2}-CD+\alpha_{1}\alpha_{2}\sqrt{(k_{\mathrm{r}}^{2}-k_{\mathrm{r},y}^{2})}\sqrt{(k_{\mathrm{t}}^{2}-k_{\mathrm{r},y}^{2})}}\right).\label{Eq:phase_t}
\end{equation}
\end{widetext}

According to Refs.~\citep{Artmann48,Dadoenkova2012}, there is the
following relation between the phase shifts in reflection and transmission
and the respective GH shifts: 
\begin{equation}
\Delta_{\mathrm{r(t)}}=-\frac{\partial\varphi_{\mathrm{r(t)}}}{\partial k_{\mathrm{r(t)},y}},\label{Eq:GH}
\end{equation}
where the tangential components of the reflected and transmitted wavevectors
are equal, $k_{\mathrm{r},y}=k_{\mathrm{t},y}$. The final formulas
for the GH shifts are presented in Appendix A.

Below we assume FM-1 and FM-2 to be the same material. Thus, in Eq.
(\ref{Eq:transm}) $\theta_{\mathrm{i}}=\theta_{\mathrm{t}}$, which
implies~$T=t^{2}$.

\subsection{Micromagnetic simulations}

Micromagnetic simulations (MSs) are a proven and efficient tool for
the calculation of the SW dynamics in various geometries \citep{Venkat13,Kim11,Hertel04,Lebecki08}.
We have used the GPU\nobreakdash-accelerated MS program MuMax3~\citep{Vansteenkiste14},
which solves the time-dependent LL equation with a Landau damping
term by the finite difference method. 
\begin{figure}
\includegraphics[width=8.6cm]{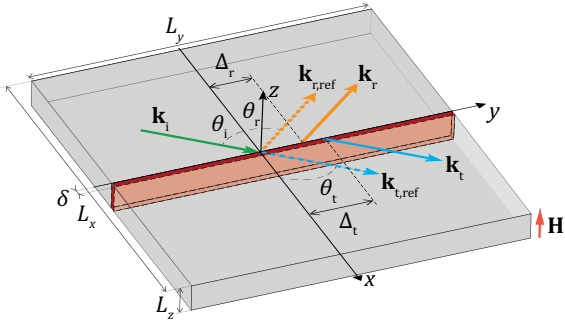} \protect\caption{Schematic representation of the simulated system. The structure is
a thin film with a thickness $L_{z}$ much smaller than its lateral
dimensions $L_{x}$ and $L_{y}$. The red area at $y=0$ is an interface
layer with a width $\delta$; $\mathbf{k_{\mathrm{i}}}$, $\mathbf{k_{\mathrm{r}}}$
and $\mathbf{k_{\mathrm{t}}}$ are the wavevectors of incident, reflected
and transmitted SW beams, respectively; the wavevectors of GH shift-free
reference reflected and refracted beams are denoted as $\mathbf{k_{\mathrm{r,ref}}}\mbox{ and }\mathbf{k_{\mathrm{t,ref}}}$,
respectively; $\Delta_{\mathrm{t}}$ is the total lateral shift (along
the interface) of the transmitted SW beam with respect to the incident
beam. \label{fig:fig1}}
\end{figure}

Shown in Fig.~\ref{fig:fig1}, the simulated system consists of two
extended Py thin films with dimensions $L_{x}/2\times L_{y}\times L_{z}$,
separated by a narrow interface slice with a width %
\mbox{%
$\delta=2$~nm%
}. The interface is parallel to the $yOz$ plane and has a magnetic
anisotropy different from that in Py. The surface magnetic anisotropy
$K_{\mathrm{S}}$ at the interface between the Py films is introduced
in the MSs by assuming a uniaxial magnetic anisotropy value $K_{12}$
in the interface slice. Its all other parameters are the same as elsewhere
in this magnetic system, comprising the films and the interface region.

We assumed a Py saturation magnetization $M_{1}=M_{2}=M_{\mathrm{S}}=0.7\times10^{3}$
Gs, and an exchange constant $A_{1}=A_{2}=1.1\times10^{-6}$ erg/cm.
The simulations were performed for a magnetically saturated film with
a thickness $L_{z}=10$~nm. A magnetic field $H_{0}=15$ kOe was
applied along the $z$-axis. The structure was discretized into cuboid
elements with in-plane dimensions 2~nm~$\times$~2~nm, much smaller
than both the exchange length of Py (6~nm) and the wavelength of
the SWs. In the $z$ direction each cuboid extended across the thickness
of the film. To speed up the simulations, the lateral dimensions $L_{x}\times L_{y}$
of the simulated area were assumed depending on the angle of incidence.
Nonetheless, in all the cases considered these dimensions were large
enough (several $\mu$m) to prevent the influence of the finite size
of the system on the propagation of SWs. Moreover, absorbing boundary
conditions were assumed in order to prevent SW reflection by the edges
of the simulated structure~\citep{Gruszecki14}

The MSs were performed in two stages, static and dynamic. In the first,
static stage an equilibrium magnetic configuration of the system was
obtained by relaxing a random magnetic configuration in the presence
of strong damping ($\alpha=0.5$). In the second, dynamic stage of
the simulations, the static configuration was slightly perturbed by
a harmonic dynamic external magnetic field $\mathbf{H}_{\mathrm{dyn}}(x,y,t)=H_{x}(x,y)\sin\left(2\pi ft\right)\hat{e}_{x}$
(i.e., $\mathbf{H}\perp\mathbf{H}_{\mathrm{dyn}}$) to induce continuous
SW excitation. The frequency $f$ of the field $\mathbf{H}_{\mathrm{dyn}}$
determines the frequency of the excited SWs. The spatial profile of
this field, $H_{x}(x,y)$, and its correspondence to the SW dispersion
determine the shape of the SW excitation and its direction of propagation
\citep{Gruszecki15-antennas}. The shape and amplitude of the dynamic
field were designed to excite a Gaussian SW beam (see Appendix~B
for a detailed description of the SW beam generation procedure). The
generated Gaussian SW beam was $500$~nm wide in its waist and had
a frequency~$f=100$~GHz. After sufficiently long continuous SW
excitation, when the transmitted beam became clearly visible, data
necessary for further analysis were acquired and saved. In the dynamic
simulations we assumed a reduced finite value of the damping parameter,
$\alpha=0.0005$, \cite{footnote2} to ensure long-distance propagation
of the SW beam.

\begin{figure}
\includegraphics[width=8.6cm]{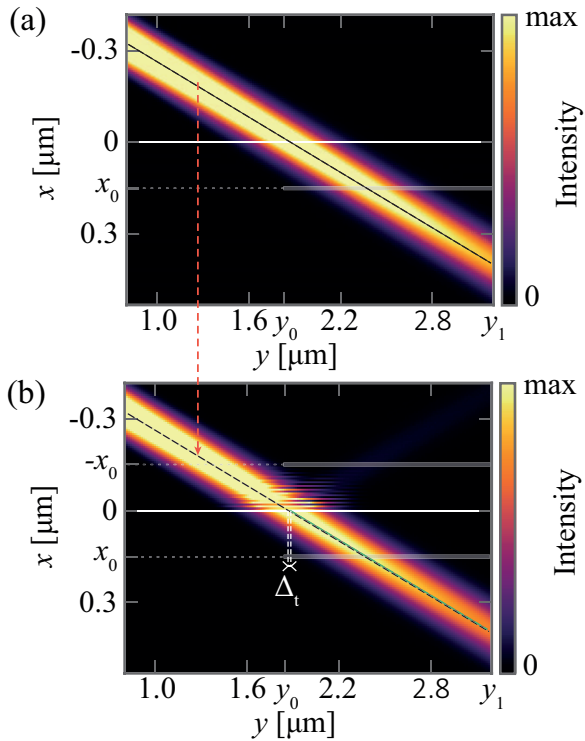} \protect\caption{Sample results of the micromagnetic simulations for $\theta_{i}=60^{\circ}$,
with intensity maps obtained from (a) the reference simulation of
SW propagation in a homogeneous Py film, and (b) simulations of SWs
transmitted through an interface with a strong magnetic anisotropy,
$K_{12}=-4.5\times10^{6}$ erg/cm$^{3}$. The green dashed line in
(b) is the ray of the refracted SW beam; the black dashed line is
the ray of the reference beam, taken from (a). The horizontal white
solid line represents the interface, and the horizontal gray solid
lines the intervals used for calculating the intensities of the transmitted
and reflected SW beams. The extracted value of the GH shift of the
transmitted beam is $\Delta_{\mathrm{t}}=3.1$~nm.\label{fig:fig2}}
\end{figure}

The acquired simulation data were then processed in order to extract
the reflectance~$R$ and transmittance~$T$, and the GH shift~$\Delta_{\mathrm{t}}$
of the transmitted beam. Firstly, time-average SW intensity colormaps
(SWICs) were drawn for all the simulation results, with the time-average
SW intensity~ $I(x,y)$ calculated from the equation: 
\[
I(x,y)=\frac{f}{4}\int_{0}^{4/f}\left[m_{x}\left(x,y,t\right)\right]^{2}\mathrm{d}t.
\]
Then, the value of $T$ was extracted as the ratio of the SW power
flowing through a plane parallel to the interface and shifted from
it by a distance~$x_{0}$ and the SW power in reference simulation:
\[
T=\left[\int_{y_{0}}^{y_{1}}I(x_{0},y)\mathrm{d}y\right]/\left[\int_{y_{0}}^{y_{1}}I_{\mathrm{ref}}(x_{0},y)\mathrm{d}y\right];
\]
the integration limits $y_{0}$ and $y_{1}$ along the $y$-axis are
indicated in Fig.~\ref{fig:fig2}(a) and (b) (bold gray lines), where
they play the role of numerator and denominator integration limits,
respectively. The reference simulation SWIC~$I_{\mathrm{ref}}(x,y)$
was obtained from simulations performed for a uniform Py thin film
without any perturbation in the interface area.

The reflectance $R$ can be calculated in a similar way, with the
intensity of SWs flowing through a surface located at the position
$-x_{0}$, normalized with the reference simulation result at $x_{0}$:
\[
R=\left[\int_{y_{0}}^{y_{1}}I(-x_{0},y)\mathrm{d}y\right]/\left[\int_{y_{0}}^{y_{1}}I_{\mathrm{ref}}(x_{0},y)\mathrm{d}y\right].
\]
We used this approach to discard the influence of the finite damping
in the MSs. The relation $R+T=1$ is fulfilled with good accuracy
by our simulation results.

A similar approach was used for extracting the GH shift from the simulation
data~\citep{Gruszecki15}. By Gaussian fitting we extracted the positions
of the centers of the intensity profiles~$I(x,y_{i})$ along many
lines (over 100 of points~$i$) perpendicular to the interface at
$y=y_{i}$. Then, having the coordinates of the centers of the incident
and transmitted beams, we calculated the coefficients of the straight
lines corresponding to the beams (dashed black line in Fig.~\ref{fig:fig2}(a)
and (b) for incident beam and dashed green line in Fig.~\ref{fig:fig2}(b)
for transmitted beam). Finally, we calculated the GH shift~$\Delta_{\mathrm{t}}$,
i.e., the horizontal shift between the transmitted and reference beams.
Because of the weak reflectance and an ambiguity in the choice of
the reflection plane in the numerical extraction of the reference
beam we do not estimate here the GH shift~$\Delta_{\mathrm{r}}$
of the reflected beam. It is noteworthy that the value of~$\Delta_{\mathrm{t}}$
is free of such ambiguity, since it does not depend on the choice
of the interface position. 

\section{Results \label{sec:Results}}

\subsection{Analytical model}

\begin{figure}
\includegraphics[width=8.6cm]{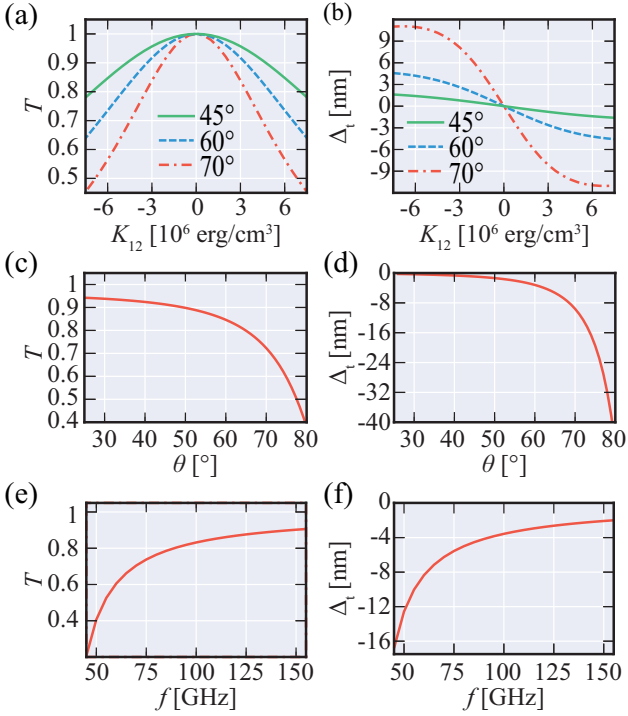} \protect\caption{(a) Transmittance and (b) GH shift vs. magnetic anisotropy at the
interface between two semi-infinite materials~(Py). Green solid,
blue dashed and red dash-dotted lines correspond to different angles
of incidence $\theta_{\text{i}}$: $45^{\circ}$, $60^{\circ}$ and
$70^{\circ}$, respectively. (c) Transmittance and (d) GH shift vs.
angle of incidence for $K_{12}=4.5$ erg/cm$^{3}$. Plots (a)-(d)
were obtained for SWs of frequency $f=100$~GHz. (e) Transmittance
and (f) GH shift vs. frequency for $K_{12}=4.5$ erg/cm$^{3}$ and
$\theta_{\mathrm{i}}=60{}^{\circ}$. The results presented in this
Figure were obtained for $H_{0}=15$~kOe.\label{fig:Results-dX-theory}}
\end{figure}

Let us first analyze the results obtained from the analytical model
for high-frequency SWs in a structure (schematically depicted in Fig.~\ref{fig:FigAM_schema})
composed of two semi-infinite Py films separated by an interface with
a width much smaller than the wavelength of the SWs ($\delta\ll\lambda$).
In the calculation of the effective material parameters in the interface
region we assumed the same value of the interface width as in the
micromagnetic simulations, $\delta=2$~nm. The transmittance defined
in Eqs.~(\ref{Eq:transm}) and (\ref{Eq:t}) is a symmetric function
of $K_{12}$; obviously, for $K_{12}=0$ (which corresponds to a homogeneous
medium) reflection does not occur and $T=1$. At a fixed angle of
incidence the value of $T$ decreases with increasing $\left|K_{12}\right|$
(see Fig.~\ref{fig:Results-dX-theory}(a)); $T$ decreases also with
increasing angle of incidence (Fig.~\ref{fig:Results-dX-theory}(c)).
A similar behavior is observed in electromagnetic waves. However,
in the case considered, at a fixed angle of incidence the transmission
of SWs through an interface with an increased magnetic anisotropy
increases with increasing SW frequency (Fig.~\ref{fig:Results-dX-theory}(e))
\citep{Gorobets98}.

In the case of identical materials with zero anisotropy ($K_{1}=K_{2}=0$)
the GH shifts for reflected and refracted SWs are equal, $\Delta_{\mathrm{t}}=\Delta_{\mathrm{r}}$.
This results from identical expressions for the phase shifts $\varphi_{\mathrm{r}}$
and $\varphi_{\mathrm{t}}$, see Eqs.~(\ref{Eq:phase_r}) and (\ref{Eq:phase_t}).
From those equations it also follows that the dependence of the GH
shift on the magnetic anisotropy at the interface is an antisymmetric
function of~$K_{12}$, taking on positive or negative values for
$K_{12}<0$ and $K_{12}>0$, respectively (see Fig.~\ref{fig:Results-dX-theory}(b)).
This is caused by the change from easy-axis to easy-plane magnetic
anisotropy upon reversal of the sign of $K_{\text{12}}$ \citep{Gruszecki14}.

The absolute value of the GH shift, ~$|\Delta_{\mathrm{t}}|$, increases
with increasing angle of incidence for $|K_{12}|\leq6\times10^{6}$
erg/cm$^{3}$ (see Fig.~\ref{fig:Results-dX-theory}(b) and (d)).
As in other types of waves, the GH shift increases substantially with
the angle of incidence, up to values comparable to the wavelength
of the SWs (around 60~nm at 100~GHz). However, for absolute values~$|K_{12}|$
of the magnetic anisotropy larger than those considered here $\left|\Delta_{\mathrm{t}}\right|$
decreases to approach~0 in the limit $|K_{12}|\rightarrow\infty$.
This dependence is similar to that observed in reflection by the edge
of a ferromagnetic material with a surface anisotropy~\citep{Gruszecki14}.

Another observation is a decrease in the GHS with increasing frequency,
shown in Fig.~\ref{fig:Results-dX-theory}(f). This implies small
phase shifts in transmission of high-frequency SWs. In general, analysis
of the plots in Fig.~\ref{fig:Results-dX-theory} leads to the conclusion
that the GH shift increases with decreasing transmission. However,
this only applies to a limited range of interface anisotropy constant
values, i.e., until $|\Delta_{\mathrm{t}}|$ reaches a maximum.

\subsection{Simulations}

The results of the analytical modeling were obtained for exchange
SWs uniform along the $z$-axis, propagating in an infinitely thick
material (Py). The micromagnetic simulations, however, take into account
the dipolar interaction and the finite thickness of the sample; thus,
in the MSs we investigate reflection and transmission of SWs in a
thin-film system of finite thickness, consisting of two films\inputencoding{latin1}{
}\inputencoding{latin9}connected by an edge interface with a width
$\delta=2$~nm (Fig.~\ref{fig:fig2}). The considered range of $K_{12}$
values in the interface region is $\left|K_{12}\right|\leq4.5\times10^{6}$
erg/cm$^{3}$, a physically realistic magnetic anisotropy~\citep{Vaz2008}.
Note that in pure Py the volume anisotropy is usually negligible.\inputencoding{latin1}{ }

\begin{figure}
\includegraphics[width=8.6cm]{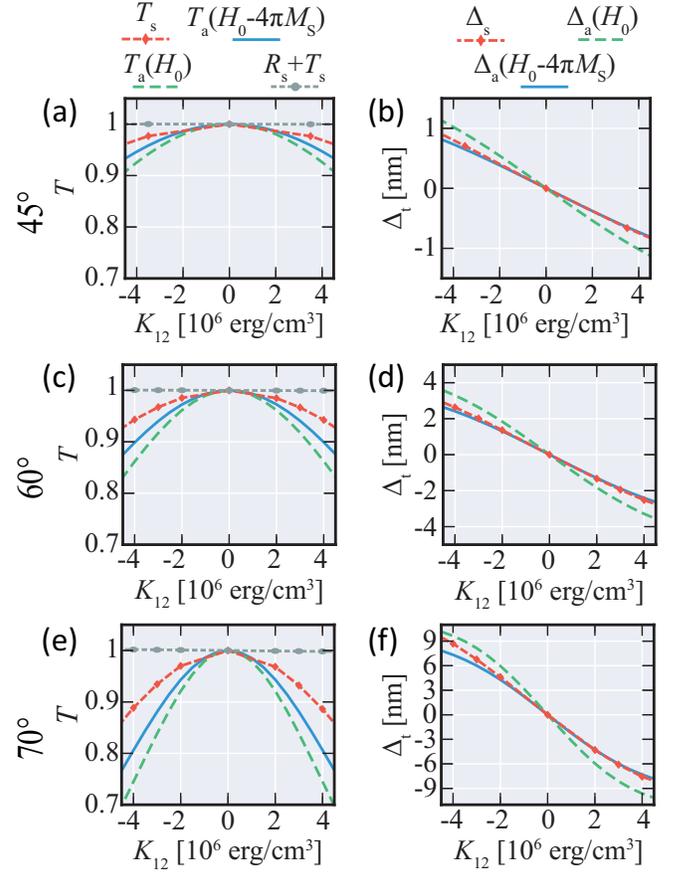} \protect\caption{Analytical results and simulation data. The MS results (red dashed
lines with diamonds) were obtained for a Py thin film divided by an
interface with anisotropy~$K_{12}$. In the analytical model we considered
two semi-infinite materials (Py) separated by an interface plane with
anisotropy~$K_{12}$, with internal magnetic field $H_{0}$ (green
dashed lines), later artificially reduced to $H_{0}-4\pi M_{\mathrm{S}}$
(blue solid line). (a),~(c),~(e)~Transmittance versus $K_{12}$
for angles of incidence $45\text{\textdegree}$, $60\text{\textdegree}$
and $70\text{\textdegree}$, respectively. Plotted as a dashed gray
line with circles, the sum of the MS values of transmittance and reflectance
is in very good approximation equal to~1. (b),~(d),~(f)~Goos-H%
\mbox{%
ä%
}nchen shift $\Delta_{\mathrm{t}}$ vs.~$K_{12}$ for angles of incidence
$45^{\circ}$, $60^{\circ}$ and $70^{\circ}$, respectively.\label{fig:Results-dX}}
\end{figure}

\inputencoding{latin9}Figure~\ref{fig:Results-dX} shows $T$ (left
column) and $\Delta_{\mathrm{t}}$ (right column) plotted vs. $K_{12}$
for three angles of incidence: 45\textdegree , 60\textdegree{} and
70\textdegree . The simulation results are represented by red dashed
lines and labeled with a subscript~$s$. These results are qualitatively
consistent with those obtained from the analytical model (green dashed
lines and subscript~$a$ in Fig.~\ref{fig:Results-dX}). However,
the transmittance values are higher and the GH shifts smaller than
the respective analytical results.

We suppose this is due to the finite thickness of the Py film and
the dipolar interaction in the MSs. In the perpendicular configuration
of the magnetization with respect to the film plane the dipolar interaction
creates a strong static demagnetizing field, which reduces the internal
magnetic field by~$4\pi M_{\mathrm{S}}$. Indeed, the substitution
of a reduced value of the external magnetic field $H_{0}^{'}=H_{0}-4\pi M_{\mathrm{S}}$
for $H_{0}$ in the analytical formulas (\ref{Eq:t}) and (\ref{Eq:GH})
for $T$ and $\Delta_{\mathrm{t}}$ provides a far better match between
the analytical results and the simulation data (see Fig.~\ref{fig:Results-dX},
where the updated analytical results are plotted as solid blue lines).

However, the MS values of transmittance are still higher than those
obtained from the analytical model. The difference increases with
the angle of incidence and with~$|K_{12}|$. We think this increase
in transmission may result from the dynamic magnetostatic interaction,
which is neglected in the model.

The GH shift obtained from Eq.~(\ref{Eq:GH}) with the reduced magnetic
field is in very good agreement with the MS results. Still, there
are some discrepancies for large angles of incidence and negative
values of~$K_{12}$ (see Fig.~\ref{fig:Results-dX}(f) for $\theta_{\mathrm{i}}=70\text{\textdegree}$
and $K_{12}<0$). Below we address this asymmetry observed in the~$\Delta_{\mathrm{t}}(K_{12})$
function.

%For all the considered angles of incidence and values of $K_{12}$ the transmission is higher than 75\%.

\begin{figure}
\includegraphics[width=8.6cm]{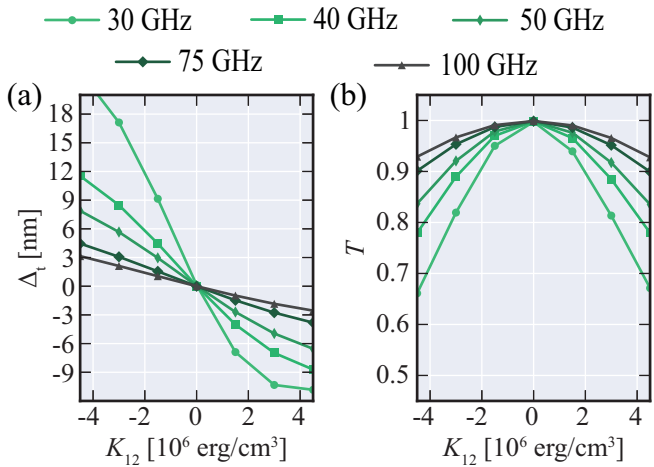} \protect\caption{Results of MSs for a Py film divided by an interface with anisotropy~$K_{12}$:
(a) $\Delta_{\mathrm{t}}$ and (b) $T$ versus $K_{12}$ for five
SW frequencies, 30~GHz, 40~GHz, 50~GHz, 75~GHz and 100~GHz. The
angle of incidence is $60\text{\textdegree}$.\label{fig:Results-dX_of_fq}}
\end{figure}

Let us analyze how the transmission of SWs changes with decreasing
SW frequency in this context. Figure~\ref{fig:Results-dX_of_fq}
shows the results of MSs with the same structure as above, for $\theta_{\mathrm{i}}=60^{\circ}$
and five different frequencies: 100~GHz, 75~GHz, 50~GHz, 40~GHz
and 30~GHz. As predicted analytically (Fig.~\ref{fig:Results-dX}),
the transmittance decreases and the absolute value of the GH shift
increases with decreasing SW frequency. However, as the SW frequency
decreases, the $\Delta_{\mathrm{t}}(K_{12})$ function becomes increasingly
asymmetric: for negative values of $K_{12}$ the absolute value of
the GH shift is larger than for positive anisotropy values (see, e.g.,
the dependence for 30~GHz, represented by a line with full dots in
Fig.~\ref{fig:Results-dX_of_fq}(a), or the plot in Fig.~\ref{fig:Results-dX}(f)). 

The analytical model was developed for purely exchange SWs, and further
extended by including the static demagnetizing field. This approximation
is well suited to high-frequency SWs. However, with decreasing SW
frequency the role of the dynamic dipolar interaction increases, and
larger discrepancies between analytical results and simulation data
can be expected.

We attribute the observed differences in transmittance to the dynamic
dipolar interaction (at %
\mbox{%
$K_{12}=-4.5\times10^{6}$%
} the discrepancy between the analytical predictions and the MS results
grows from 0.05 for 100~GHz to 0.1 for 50~GHz). However, the influence
of the dipolar interaction on SW transmission has not been fully elucidated
to date; this requires the development of a new model, which goes
beyond the scope of this paper. 

Nevertheless, the dipolar interaction does not provide explanation
of the asymmetry, also increasing with decreasing SW frequency in
the MSs, in the $K_{12}$ dependence of the GH shift. We suppose it
is due to different SW polarizations at the interface for positive
and negative~$K_{12}$ (easy axis and easy plane, respectively).
In the case of $K_{12}<0$ an increase of the $m_{y}$ component with
respect to $m_{x}$ can be expected (and is confirmed by MSs), resulting
in a larger GH shift along the $y$-axis. Indeed, whereas for $K_{12}>0$
the GH shift obtained in the simulations is in good agreement with
the analytical results (with circular precession of the magnetization
assumed), as shown in Fig.~\ref{fig:Results-dX}(f), for $K_{12}<0$
the MS values of $\Delta_{\mathrm{t}}$ are larger and the difference
increases with decreasing $K_{12}$.

\section{Summary\label{sec:Discussion}}

We have studied theoretically the reflection and transmission of obliquely
incident SWs by a thin interface between two extended ferromagnetic
media and between two semi-infinite films connected by an edge. The
interface area has a different uniaxial anisotropy with respect to
the extended areas. We have derived analytical formulas describing
the reflectance and transmittance of exchange SWs, as well as the
phase shifts and lateral GH shifts for refracted and reflected SWs.
Moreover, using micromagnetic simulations we have demonstrated that
these results of analytical modeling also describe qualitatively SWs
in the thin-film geometry. It is noteworthy that, in spite of a number
of assumptions in the analytical approach (negligible dipolar interaction,
infinite extent of the ferromagnetic materials, and circular precession
of SWs), the results obtained by the two techniques are in very good
agreement for high-frequency SWs.

In the numerical study we have focused on the transmission of SWs
and the related GH shift. The GH shift of transmitted SWs proves to
increase with increasing anisotropy in the interface region and with
decreasing SW frequency. We point out the increased role of the dynamic
dipolar interaction in SW transmission, and the influence of the elliptical
polarization of the SW beam at the interface on the GH shift at lower
frequencies. The influence of these two factors on the GH shift of
transmitted SWs needs to be elucidated in detail, which requires further
investigation and the development of a new analytical model.

The demonstrated lateral GH shift of SWs refracted by an interface
with a width much smaller than the wavelength of the considered waves
points to the possibility of steering SWs in thin films at sub-wavelength
distances. Although the lateral shift found in a Py thin film divided
in two parts by a narrow (2 nm wide) interface with increased magnetic
anisotropy is not significant, we have shown possible ways to increase
it. Further investigation can lead to the development of more methods
of phase modulation at sub-wavelength distances. This sets a promising
direction in the study of magnonic metasurfaces, a novel field in
magnonics, next to the graded index magnonics \citep{Davies15}.

In the present paper we have also proposed an efficient method for
the excitation of SW beams, which should be of use in further numerical
investigations.

\section*{Appendix A}
\begin{widetext}
The formula for the lateral GH shift of SWs reflected by an interface
between two ferromagnetic materials reads: 
\begin{equation}
\begin{array}{c}
\Delta_{\mathrm{r}}=-\dfrac{\partial\varphi_{\mathrm{r}}}{\partial k_{\mathrm{r},y}}=\\
=-\dfrac{\dfrac{\alpha_{1}k_{\mathrm{r},y}}{\sqrt{k_{1}^{2}-k_{\mathrm{r},y}^{2}}}\left(D\alpha_{2}^{2}\left(k_{\mathrm{t}}^{2}-k_{\mathrm{r},y}^{2}\right)-C\left(A_{12}^{2}-CD\right)\right)+\dfrac{\alpha_{2}k_{\mathrm{r},y}}{\sqrt{k_{\mathrm{t}}^{2}-k_{\mathrm{r},y}^{2}}}\left(C\alpha_{1}^{2}\left(k_{\mathrm{r}}^{2}-k_{\mathrm{r},y}^{2}\right)-D\left(A_{12}^{2}-CD\right)\right)}{\left(A_{12}^{2}-CD+\alpha_{1}\alpha_{2}\sqrt{k_{\mathrm{r}}^{2}-k_{\mathrm{r},y}^{2}}\sqrt{k_{\mathrm{t}}^{2}-k_{\mathrm{r},y}^{2}}\right)^{2}+\left(C\alpha_{1}\sqrt{k_{\mathrm{r}}^{2}-k_{\mathrm{r},y}^{2}}+D\alpha_{2}\sqrt{k_{\mathrm{t}}^{2}-k_{\mathrm{r},y}^{2}}\right)^{2}}\\
-\dfrac{\dfrac{\alpha_{1}k_{\mathrm{r},y}^{2}}{\sqrt{k_{\mathrm{r}}^{2}-k_{\mathrm{r},y}^{2}}}\left(D\alpha_{2}^{2}\left(k_{\mathrm{t}}^{2}-k_{\mathrm{r},y}^{2}\right)-C\left(A_{12}^{2}-CD\right)\right)-\dfrac{\alpha_{2}k_{\mathrm{r},y}}{\sqrt{k_{\mathrm{t}}^{2}-k_{\mathrm{r},y}^{2}}}\left(C\alpha_{1}^{2}\left(k_{\mathrm{r}}^{2}-k_{\mathrm{r}\parallel}^{2}\right)-D\left(A_{12}^{2}-CD\right)\right)}{\left(A_{12}^{2}-CD-\alpha_{1}\alpha_{2}\sqrt{k_{\mathrm{r}}^{2}-k_{\mathrm{r},y}^{2}}\sqrt{k_{\mathrm{t}}^{2}-k_{\mathrm{r},y}^{2}}\right)^{2}+\left(C\alpha_{1}\sqrt{k_{\mathrm{r}}^{2}-k_{\mathrm{r},y}^{2}}-D\alpha_{2}\sqrt{k_{\mathrm{t}}^{2}-k_{\mathrm{r},y}^{2}}\right)^{2}}.
\end{array}
\end{equation}

For SWs transmitted through the interface the GH shift is expressed
by the equation: 
\begin{equation}
\begin{array}{c}
\Delta_{\mathrm{t}}=-\dfrac{\partial\varphi_{\mathrm{t}}}{\partial k_{\mathrm{t},y}}=\\
=-\dfrac{\dfrac{\alpha_{1}k_{\mathrm{r},y}}{\sqrt{k_{\mathrm{r}}^{2}-k_{\mathrm{r},y}^{2}}}\left(D\alpha_{2}^{2}\left(k_{\mathrm{t}}^{2}-k_{\mathrm{r},y}^{2}\right)-C\left(A_{12}^{2}-CD\right)\right)+\dfrac{\alpha_{2}k_{\mathrm{r},y}}{\sqrt{k_{\mathrm{t}}^{2}-k_{\mathrm{r},y}^{2}}}\left(C\alpha_{1}^{2}\left(k_{\mathrm{r}}^{2}-k_{\mathrm{r},y}^{2}\right)-D\left(A_{12}^{2}-CD\right)\right)}{\left(A_{12}^{2}-CD+\alpha_{1}\alpha_{2}\sqrt{k_{\mathrm{r}}^{2}-k_{\mathrm{r},y}^{2}}\sqrt{k_{\mathrm{t}}^{2}-k_{\mathrm{r},y}^{2}}\right)^{2}+\left(C\alpha_{1}\sqrt{k_{\mathrm{r}}^{2}-k_{\mathrm{r},y}^{2}}+D\alpha_{2}\sqrt{k_{\mathrm{t}}^{2}-k_{\mathrm{r},y}^{2}}\right)^{2}}.
\end{array}
\end{equation}
\end{widetext}

\section*{Appendix B: Spin-wave beam excitation method for micromagnetic simulations\label{sec:Appendix-SWB_excitation}}

In our previous papers~\citep{Gruszecki14,Gruszecki15} we used a
simple method for the excitation of a SW beam with a wavevector parallel
to the $x$-axis, based on the application of an external microwave
magnetic field in the form $H_{\mathrm{dyn}}\left(x,y\right)=h\Theta_{\mathrm{H}}(x+w/2)\Theta_{\mathrm{H}}(-x+w/2)G(y)$,
where $\Theta_{\mathrm{H}}$ is the Heaviside step function, $G(y)$
is the Gaussian distribution function $G(x)=\exp\left[2x^{2}/\left(L\sigma\right)^{2}\right]$,
$w$ denotes length, $L$ is the width of the excitation area, and
$\sigma\le0.2$. This method is very simple to implement in MSs and
provides good quality beams in most cases. However, it is far from
experimental realization and involves the excitation of some additional,
undesirable waves due to the finite discretization used in the finite
difference method. Usually this disadvantage can be neglected, but
in our study, which required very high accuracy for the determination
of the GH shifts, the additional waves represented a substantial interference.
Therefore we have developed an advanced SW beam excitation method,
in which numerical artifacts due to discretization of the applied
magnetic field are almost entirely eliminated. Moreover, the profiles
of the applied field are more realistic~\citep{Gruszecki15-antennas}
than those used previously~\citep{Gruszecki14,Gruszecki15}.

\begin{figure}
\includegraphics[width=8.6cm]{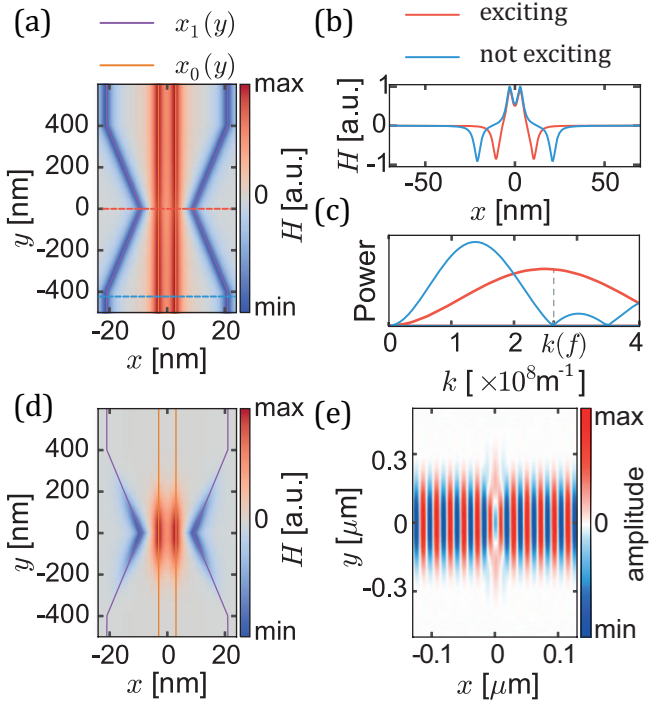} \protect\caption{(a) Profile of dynamic magnetic field $H_{\mathrm{dyn}}$ for $B(y)=1$.
(b) Magnetic field profiles along the $x$-axis: $H_{1}=H_{x}(y=0)$
(red line) and $H_{2}=H_{x}(y=L/2)$ (blue line). (c) Fourier transforms
of the magnetic field profiles presented in (b): $\widetilde{H}_{1}=\mathcal{F}\left\{ H_{1}\right\} $
(red) and $\widetilde{H}_{2}=\mathcal{F}\left\{ H_{2}\right\} $ (blue).
Note that $\widetilde{H}_{1}$ has a maximum for $k=k(f)$ corresponding
to a zero of $\widetilde{H}_{2}$. (d) Profile of dynamic magnetic
field $H_{\mathrm{dyn}}$ for $B(y)=G(y)$. (e) Sample SW beam excited
using the dynamic magnetic field profile (d).\label{fig:figA1}}
\end{figure}

Let us assume a dynamic field in the form: 
\begin{equation}
H_{x}(x,y,t)=A(x,y)B(y)\sin\left(2\pi ft\right),
\end{equation}
where the function $B(y)$ describes the envelope of the magnetic
field amplitude along the $y$-axis. Usually, especially in the case
of SW excitation in waveguides or in homogeneous structures, we can
assume~$B(y)=1$. However, sometimes, particularly for oblique SW
beam excitation in planar structures, other envelopes should be used,
which we will present later.

The function $A(x,y)$ approximates the profile of the field generated
by a coplanar waveguide antenna~\citep{Gruszecki15-antennas}: 
\begin{equation}
\begin{array}{cc}
A(x,y)=C & \left[\frac{1}{\left(x+x_{1}\left(y\right)\right)^{2}+1}-\frac{1}{\left(x+x_{0}\left(y\right)\right)^{2}+1}\right.\\
 & \left.-\frac{1}{\left(x-x_{0}\left(y\right)\right)^{2}+1}+\frac{1}{\left(x-x_{1}\left(y\right)\right)^{2}+1}\right],
\end{array}
\end{equation}
where $C$ is a constant, the value of which determines the maximal
value of~$H_{x}$.

The functions $x_{0}\left(y\right)$ and $x_{1}\left(y\right)$ describe
the geometry of the generated field and determine the efficiency of
SW excitation. According to an analysis presented in Ref.~\citep{Gruszecki15-antennas},
for such a profile the maximally efficient resonant excitation of
SWs with a wavelength $\lambda$ is achieved when $x_{0}(0)\mbox{ and }x_{1}(0)$
fulfill the equation: 
\begin{equation}
\frac{2}{\lambda}=\frac{1}{x_{0}(0)+x_{1}(0)},
\end{equation}
where $\lambda$ is a function of frequency, $\lambda(f)=2\pi/k(f)$,
via the inverse dispersion relation~$k(f)$.

In the case of exchange-dominated SWs Eq.~(\ref{Eq:Dispersion})
can be used; the general dispersion relation is presented in Ref.~\citep{Kalinikos86}.

On the other hand, with the geometrical parameters $x_{0}(y_{\mathrm{nE}})$
and $x_{1}(y_{\mathrm{nE}})$ fulfilling the equation: 
\begin{equation}
\frac{1}{\lambda}=\frac{1}{x_{0}(y_{\mathrm{nE}})+x_{1}(y_{\mathrm{nE}})},
\end{equation}
the excitation of SWs will be inefficient. The solution of this set
of equations leads to: 
\begin{equation}
2x_{0}(0)+2x_{1}(0)=x_{0}(y_{\mathrm{nE}})+x_{1}(y_{\mathrm{nE}}).
\end{equation}
For simplicity, let us assume a constant value of %
\mbox{%
$x_{0}(y)\equiv x_{0}$%
}. Thus: 
\begin{equation}
x_{1}(y_{\mathrm{nE}})=x_{0}+2x_{1}(0).
\end{equation}

We assume $x_{1}(y)$ is a continuous linear function: 
\begin{equation}
x_{1}(y)=x_{1}(0)+\frac{2}{L}\left[x_{0}+x_{1}(0)\right]y,
\end{equation}
where $L/2=y_{\mathrm{nE}}$ is the length over which the value of
$x_{1}(y)$ changes from $x_{1}(0)$ (corresponding to resonant SW
excitation) to $x_{1}(y_{\mathrm{nE}})$ (corresponding to inefficient
SW excitation). Note that $L$ can be regarded as the length of the
antenna; for $y>L/2$ SWs will not be excited.

This approach only applies to the case $L\gg\lambda$. Moreover, the
substitution of $G(y)$ for $B(y)$ is recommended for further improvement
of the quality of the simulated beam. Alternatively, $\mathrm{sech}^{2}$
can be used instead of the Gaussian function as an additional envelope
of the dynamic magnetic field.

In the MSs presented in this paper $500$~nm wide SW beams were excited
with the following parameter values: $L=1200$~nm, $\sigma^{2}=0.1$
and $C=0.02H_{0}$.
\begin{acknowledgments}
This project has received funding from the European Union's Horizon
2020 research and innovation programme under Marie Sk\l odowska-Curie
grant agreement no. 644348, and from the Polish National Science Centre,
project UMO-2012/07/E/ST3/00538. P.G. also acknowledges support %OS: financial support?
from Adam Mickiewicz University Foundation. The numerical calculations
were performed at Poznan Supercomputing and Networking Center (grant
no. 209). 
\end{acknowledgments}

\end{document}